# Extreme magnetoresistance and Shubnikov-de Haas oscillations in ferromagnetic DySb

D. D. Liang,[1,2,a] Y. J. Wang,[1,a] C. Y. Xi,[1] W. L. Zhen,[1] J. Yang,[1] L. Pi,[1,2] W. K. Zhu,[1,b] and C. J. Zhang[1,3,b]

[1]*Anhui Province Key Laboratory of Condensed Matter Physics at Extreme Conditions, High Magnetic Field Laboratory, Chinese Academy of Sciences, Hefei 230031, China*

[2]*Hefei National Laboratory for Physical Sciences at Microscale, University of Science and Technology of China, Hefei 230026, China*

[3]*Institute of Physical Science and Information Technology, Anhui University, Hefei 230601, China*

[a]These authors contributed equally to this work.

[b]wkzhu@hmfl.ac.cn, zhangcj@hmfl.ac.cn.

The electronic structures of a representative rare earth monopnictide (i.e., DySb) under high magnetic field (i.e., in the ferromagnetic state) are studied from both experimental and theoretical aspects. A non-saturated extremely large positive magnetoresistance (XMR) is observed (as large as $3.7 \times 10^4$% at 1.8 K and 38.7 T), along with the Shubnikov-de Haas oscillations that are well reproduced by our first principles calculations. Three possible origins of XMR are examined. Although a band inversion is found theoretically, suggesting that DySb might be topologically nontrivial, it is deeply underneath the Fermi level, which rules out a topological nature of the XMR. The total densities of electron-like and hole-like carriers are not fully compensated, showing that compensation is unlikely to account for the XMR. The XMR is eventually understood in terms of high mobility that is associated with the steep linear bands. This discovery is important to the intensive studies on the XMR of rare earth monopnictides.

The past two or three years have witnessed the renaissance of rare earth monopnictides,[1-15] which represent a novel electronic system for the extremely large positive magnetoresistance (XMR) and the potential to realize exotic topological semimetal (TSM) state, both of which are essential functionalities of electronic materials for magneto-electronics and fast electronics applications. XMR was first reported in classic semimetal Bi,[16-18] and TSMs like WTe$_2$,[19] Cd$_3$As$_2$,[20] and TaAs.[21] For monopnictides, the XMR of ~$10^5$% was observed in LaSb,[1] and subsequently in LaBi,[3] YSb,[4] and magnetic NdSb[5,13] and CeSb.[15] The magnetotransport in LaSb/Bi is characterized by a typical near-quadratic field dependence of MR, as well as a field-induced upturn in resistivity followed by a plateau at low temperature.[1,3] These fingerprints were also observed in WTe$_2$,[19] and ZrSiS (a nodal-line semimetal candidate),[22] which further implies a potential to realize TSM state in these systems. So far, several mechanisms have been proposed to explain the XMR of monopnictides, including electron-hole compensation,[3,6,8,12,14] forbidden backscattering at zero field,[1,13] and nontrivial band topology.[6] However, no band inversion or crossing was observed by angle-resolved photoemission spectroscopy (ARPES) measurements for LaSb or CeSb,[6,7,11] showing that they are topologically trivial semimetals. The origin of the XMR in rare earth monopnictides is still under debate.

Another feature of rare earth monopnictides is from the magnetism of (some) rare earth ions and the tunable magnetic structures, which provides an opportunity to investigate the possible correlation between topological property and magnetism. Kuroda et al. discovered cone-like structures in CeBi by ARPES.[11] Their experiments show that the occurrence of band inversion strongly depends on the strength of spin-orbit coupling (SOC) which determines the splitting of *p* orbital and hybridization with Ce $t_{2g}$ orbital. In addition, Wu et al. show that rare earth elements have an impact on the electronic structures,[10] in view of the different lattice constants and correlation strengths. Theoretical calculations also suggest a close relation between magnetic structures and band structures.[13,15] Therefore, the occurrence of TSM state in rare earth monopnictides is critical. Only a few compounds are proposed as candidates for TSMs,[2,11,13,15] and the inversion points are all below the Fermi level (~-0.4 eV).

DySb is another representative monopnictide, as it has an enhanced SOC that pushes $p_{3/2}$ orbital high and allows possible inversion between $p_{3/2}$ orbital and Dy $d$ orbital,[10] just like the situation in CeBi. This suggests that DySb is a potential magnetic TSM. Like other rare earth antimonides, DySb adopts a rocksalt type crystal structure at room temperature, namely, cubic with space group *Fm*-3*m* (225). However, at low temperature, its crystal structure changes into a tetragonal one with a shrinkage in axis *c* (*c*/*a*=0.993).[23] The space group is hence changed into *I*4/*mmm* (139). Accompanying the structural transition, it enters antiferromagnetic (AFM) ground state,[24] in which the magnetic moments ($Dy^{3+}$ ions) are aligned in type-II AFM structure,[25] i.e., opposite ferromagnetic (FM) (011) planes alternatively stacked along [011] direction (here [011] refers to tetragonal space group). The easy axis, namely the direction of moments, is along [001] axis [Fig. 1(a)]. Such an AFM ground state can be changed into an FM state [Fig. 1(c)] by magnetic field applied along [001] direction, via an intermediate phase.[25] In the intermediate phase, the moments in one of each two nearest (011) planes are rotated by 90°, to the [-110] direction [Fig. 1(b)].[25,26] Hence, the measured magnetization is exactly one half of the saturated value. This is an important feature to characterize DySb samples.

In this work, the electronic structures of DySb under high magnetic field (i.e., in the FM state) are investigated from both experimental and theoretical aspects. A non-saturated XMR is observed (as large as $3.7 \times 10^4$% at 1.8 K and 38.7 T), along with Shubnikov-de Haas (SdH) oscillations that are well reproduced by our first principles calculations. Three possible origins of XMR are examined. Although a band inversion is found theoretically, suggesting that DySb might be topologically nontrivial, it is deeply underneath the Fermi level, which rules out a topological nature of the XMR. The total densities of electron-like and hole-like carriers are not fully compensated, showing that compensation is unlikely to account for the XMR. The XMR is eventually understood in terms of high mobility that is associated with the steep linear bands.

Single crystals of DySb were grown with tin flux method. The chemical stoichiometry was confirmed by the energy dispersive spectroscopy (EDS). The crystal structure and phase purity were checked by single crystal X-ray diffraction (XRD) on a Rigaku-TTR3 X-ray diffractometer using Cu Kα radiation. Magnetic measurements were taken on a Quantum Design MPMS. Electrical measurements were taken on a Quantum Design PPMS (14 T) and on the hybrid magnet of CHMFL (up to 40 T) at Hefei. DFT calculations were performed using the WIEN2k code.[27] The generalized gradient approximation (GGA) of Perdew-Burke-Ernzerhof[28] was employed with a method of GGA+$U$+SO, where $U$ was the effective exchange-correlation potential for Dy 4$f$ orbitals (4 eV in this work) and SO represented the spin-orbit interactions. The $k$ mesh consisted of 8,000 points. Quantum oscillation frequencies were calculated using the Supercell K-space Extremal Area Finder (SKEAF) tool.[29]

The as-grown crystal has a typical rectangular shape, as shown in the inset of Fig. 2(a). Figure 2(a) presents the single crystal XRD pattern taken at room temperature, in which only (00$l$) peaks are observed, confirming high quality of single crystal. As described above, DySb adopts a cubic structure at room temperature, and at low temperature changes into a tetragonal one [Fig. 2(b)]. In order to characterize such a transition, electrical and magnetic measurements are performed. As seen in Fig. 2(c), the temperature dependent electrical resistivity ($\rho$) shows a metallic behavior, and a clear drop appears at about 10.8 K, which is consistent with the structural transition.[24] At the same temperature, the magnetic susceptibility ($\chi$) also exhibits a sharp cusp [Fig. 2(e)], which has been identified as an AFM transition. By fitting the data above 50 K to the Curie-Weiss law [red curve in Fig. 2(e)], an effective magnetic moment of 10.8 $\mu_B$ is obtained, consistent with the theoretical value $\mu_{\text{eff}} = g\sqrt{J(J+1)}\mu_B$, where $g$ is the Landé $g$ factor (4/3) and $J$ is the total angular momentum (15/2). The resultant Weiss temperature Θ=-8 K agrees with the AFM nature.

The magnetic field dependence of resistivity and magnetization is further measured, with magnetic field applied along the [001] direction. Figure 2(d) shows the magnetoresistance (MR, $(\rho(B) - \rho(0))/\rho(0)$ ) taken at 3 K, where $\rho(B)$ and $\rho(0)$ represent the resistivity measured at magnetic field $B$ and at zero field, respectively. Two distinct features can be found from the MR curve. First, a non-saturated positive MR reaches as large as 0.54×10$^4$% at 14 T, indicating that DySb is another XMR material, like the other family members in rare earth antimonides. Second, two kinks at about 2.4 T and 4.8 T are observed, which shows signatures of field induced transitions. Such two transitions are confirmed by the field dependent magnetization measurement [Fig. 2(f)]. The *M-H* curve exhibits a stepwise form with two critical fields at about 2.5 T and 4.8 T that are roughly consistent with the kinks in MR. The magnetization is saturated above 6.4 T, i.e., $\mu_{sat} = 10\ \mu_B$. Between 2.5 T and 4.2 T, an intermediate phase emerges with a moment of 5 $\mu_B$, one half of the saturated moment. All these magnetic phases and transitions are consistent with previous measurements,[25] showing that the obtained DySb crystals are well characterized.

In the present paper, we study mainly the FM phase because the XMR and quantum oscillations occur at very high field. Figure 3(a) presents the longitudinal MR of DySb taken at various temperatures with the magnetic field sweeping from 11 T to 38.7 T. The non-saturated positive XMR is again confirmed, which reaches as large as $3.7 \times 10^4$% at 1.8 K and 38.7 T, without any sign of saturation. A simple power law fit using the formula MR$\propto B^n$ gives an exponent n=1.74 [inset of Fig. 3(a)], which is smaller than n=2 as expected for a perfectly compensated metal. Such a deviation is also observed in other XMR materials, such as WTe$_2$,[30] PtSn$_4$,[31,32] and a variety of rare earth monopnictides, e.g., LaBi/Sb,[3,6] YSb,[4] NdSb[13] and LuSb.[8]

The XMR can be discussed in the regime of two-band theory,[33] expressed as

$$\rho_{xx} = \frac{(n_e\mu_e + n_h\mu_h) + (n_e\mu_e\mu_h^2 + n_h\mu_e^2\mu_h)B^2}{e[(n_e\mu_e + n_h\mu_h)^2 + (n_h - n_e)^2\mu_e^2\mu_h^2 B^2]}, \tag{1}$$

where $\rho_{xx}$ is the longitudinal resistivity, *e* is the electron charge, *n* is the carrier density, $\mu$ is the mobility, and the subscripts e and h refer to electron-type and hole-type carriers, respectively. According to the model, the condition for $\rho_{xx}$ to increase as $B^2$ is $n_e = n_h$, which means that electron-type and hole-type carriers are perfectly compensated by each other. The deviation from 2 shown above thus indicates an imbalanced situation. Another ramification from the expression is that the MR ratio strongly depends on the average mobility $\mu_{ave}$. Assuming an ideal condition, namely, $n_e = n_h$ and $\mu_{ave} = \mu_e \approx \mu_h$, the MR ratio will be obtained as MR=$(\mu_{ave}B)^2$, i.e., the Lorentz law.[34,35] Using this formula to fit the MR curve (1.8 K) in Fig. 3(a) leads to a $\mu_{ave} = 0.48 \times 10^4$

cm$^2$/Vs. Although this value is not as ultrahigh as two-dimensional electron gas or TSMs such as Cd$_3$As$_2$,[20] it is greatly larger than common magnetic metals. Such a high mobility should contribute to the XMR of DySb.

To check the carrier densities at Fermi surface, SdH oscillations are extracted from the MR curves and analyzed. As shown in Fig. 3(b), the SdH oscillation data as a function of $B^{-1}$ are obtained by subtracting the non-oscillatory background, represented by a polynomial, from the total MR data. The amplitudes exhibit a periodic behavior and gradually decrease with the decreasing field. Figure 3(c) shows the fast Fourier transformation (FFT) spectra of the SdH oscillations, which contains four main frequencies and one harmonic, i.e., 369 T (labeled α), 745 T (β), 1094 T (α$_1$), 1276 T (δ) and 1492 T (2β). Multiple frequencies clearly indicate a multi-band feature of the Fermi surface of DySb. The oscillation frequencies are proportional to extremal cross section areas of Fermi pockets, according to Onsager relation $F = (\hbar/2\pi e)A$, where $\hbar$ is the reduced Planck constant and $A$ is the extremal cross section area perpendicular to the external magnetic field. The corresponding $A$ is calculated and listed in Table I.

The oscillations can be described by Lifshitz-Kosevich (LK) formula, with the amplitude proportional to the thermal damping factor $R(T) = (\lambda m^*T/B)/\sin(\lambda m^*T/B)$, where $\lambda = 2\pi^2 k_B m_0/e\hbar$, $k_B$ is the Boltzmann constant, $m_0$ is the mass of free electron, and $m^*$ is the effective mass of carrier. For the main frequency branches α and β, Fig. 3(d) presents their amplitudes as a function of temperature. The fits using $R(T)$ results in an effective mass $m^*$=0.69$m_0$ and 0.88$m_0$ for α and β, respectively. These values are larger than those of other rare earth monopnictides,[5,8] which might be attributed to the not very pronounced SdH oscillations. Note that the average magnetic field $B_{ave}$=26.4 T is used for $B$ in the fits.

To calculate the carrier densities from the SdH oscillation frequencies, we need also the types of each frequency, i.e., electron-type or hole-type, and the shapes of Fermi surface. DFT calculations are performed by setting the spin polarization as FM. Figure 4(a) illustrates the first Brillouin zone (BZ) with the calculated Fermi surface, and Figs. 4(b) and 4(c) are the band structures without and with SOC considered, respectively, plotted along high symmetry points of BZ (for space group *I*4/*mmm*).[36] Four bands cross the Fermi level. The red ones (labeled α and $α_1$) are at the corners of BZ and centered at Z or X point. Their shape is like rectangular bipyramids and their nature is electron-like, which can be seen from the band structures in Figs. 4(b) and 4(c) or calculated by SKEAF. The rest three bands (labeled δ, β and ζ) are located at the center of BZ (i.e., Γ point) and all hole-like. The biggest δ band is like an octahedron, inside which is the octahedral β band with surface indentation and the spherical ζ band. These results are similar to those obtained for ferromagnetic NdSb and nonmagnetic LuSb.[8,13] The main difference comes from the slight tetragonal distortion that gives rise to generated Z and X points. However, since the $c$ and $a$ axes are so close, the difference between Z and X can be almost negligible.

The bands near Fermi surface are consisting of hybridized $d$ and $p$ orbitals. In rare earth monopnictides, topological nontrivial feature usually originates from the band crossing or inversion occurring at or in the vicinity of the high symmetry X point. The "touch" of bands is critical, which requires appropriate rare earth cations (with various ionic radii and correlation strengths) and anions (with different SOC strengths),[11] and for magnetic systems also depends on spin polarization.[13] Therefore, most of rare earth monopnictides have gapped bands at X, so does DySb, although the gap is tiny. However, if the $k$-path is along the high symmetry ΓX line, the bands with SOC show an inversion near X, as highlighted by the grey circle in Fig. 4(c). This inversion arises from the approaching and contact of hole (at Γ) and electron (at X) pockets, just similar to the mechanism of type-II Weyl semimetals.[37] Indeed, the inversion point has a two-fold degeneracy, suggesting that DySb in the FM state might be a magnetic Weyl semimetal. Unfortunately, the inversion is not on the Fermi level but seated about 0.34 eV below the Fermi level. So, it should be difficult to be detected by transport measurements, as the transport properties are mainly contributed by the carriers at Fermi level. And in consequence, the origin of XMR may be not linked to a topological nature (that suppresses backscattering). It probably stems from a normal mechanism, such as compensation effect or ultrahigh mobility.

The numerical information of Fermi surface, namely extremal cross-section area, effective mass and carrier type, can be extracted from the calculation results by using the program SKEAF.[29] During the extraction, the magnetic field is set along the $c$ axis (that is also the direction of ΓZ in BZ) as is applied in experiments. As seen in Fig. 4(a), five extremal cross sections are detected. Their corresponding SdH (or de Haas-van Alphen, dHvA) frequencies are listed in Table I, most of which agree well with the experimental values, except the ζ branch. The effective mass for each detected frequency is also given by contouring the energy slopes at the Fermi surface. We can find that the effective mass of α branch is smaller than the others, which could be ascribed to a more linear energy dispersion. Indeed, as shown in Fig. 4(c), the energy dispersion is linear along the short axis of the Fermi surface at zone corner (i.e., SZG within α plane) whereas it is parabolic along the long axis (i.e., ΓX within $α_1$ plane). Such a highly anisotropic electron band has been confirmed by ARPES measurements in an analogue, i.e., LaSb.[6] It is believed that a linearly dispersive band is always related with ultrahigh mobility,[20,35] which may explain the high mobility of DySb revealed by the analyses of the XMR data.

Since all the experimentally observed frequencies have been reproduced by theoretical calculations and their carrier types determined, the total densities of electron-like carriers and hole-like carriers can be computed and compared. First, the shape of Fermi surface at zone center (corner) is roughly approximated as balls (ellipsoids). The Fermi wave-vectors $k_F$ and volumes $V_F$ are then calculated accordingly (using experimental values). Finally, the density of each Fermi surface branch is calculated with the relation $n = V_F/4\pi^3$. As shown in Table I, the total density is $3.57 \times 10^{20}$ cm$^{-3}$ for electron-like carriers (triple ellipsoidal Fermi surface in one BZ) and $3.72 \times 10^{20}$ cm$^{-3}$ for hole-like carriers. That is, the situation in DySb is not the perfect compensation, which is consistent with the deviation of n=1.74 exponent from 2, shown in the power law fit of MR data [inset of Fig. 3(a)]. Thus, compensation mechanism is unlikely to serve as the main driving force of XMR. This point is further confirmed by the gated WTe$_2$ experiments,[38] in which an electrolytic gate breaks the compensation but the XMR remains. The XMR in DySb is mainly arising from the high mobility, in association with the linearly dispersive band at zone corner. Such a linear dispersion seems to be a common feature for rare earth monopnictides, so it is probably the most important origin of the XMR in these systems.

In summary, the electronic structures of DySb under high magnetic field are investigated from both experimental and theoretical aspects. A non-saturated XMR is observed, along with the SdH oscillations that are well reproduced by our first principles calculations. Although a band inversion is found theoretically, suggesting that DySb might be topologically nontrivial, it is deeply underneath the Fermi level, which rules out a topological nature of the XMR. The XMR is eventually understood in terms of high mobility that is associated with the steep linear bands.

This work was supported by the National Key R&D Program of China (Grant Nos. 2016YFA0300404 and 2017YFA0403600), the National Natural Science Foundation of China (Grant Nos. 51603207, U1532267, 11574288 and 11674327), and the Natural Science Foundation of Anhui Province (Grant No. 1708085MA08).

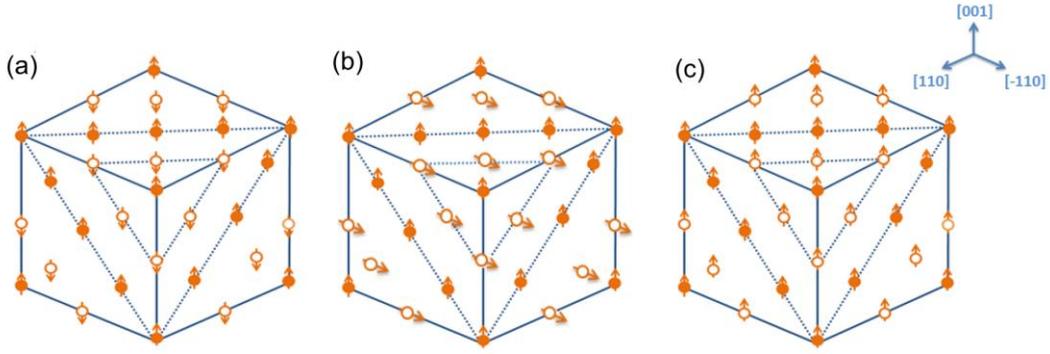

FIG. 1. Magnetic structures for (a) the AFM ground state (NiO-type), (b) the intermediate phase (HoP-type), and (c) the FM state, respectively. Only Dy atoms are presented. Coordinate axes are based on those for tetrahedral structure in Fig. 2(b).

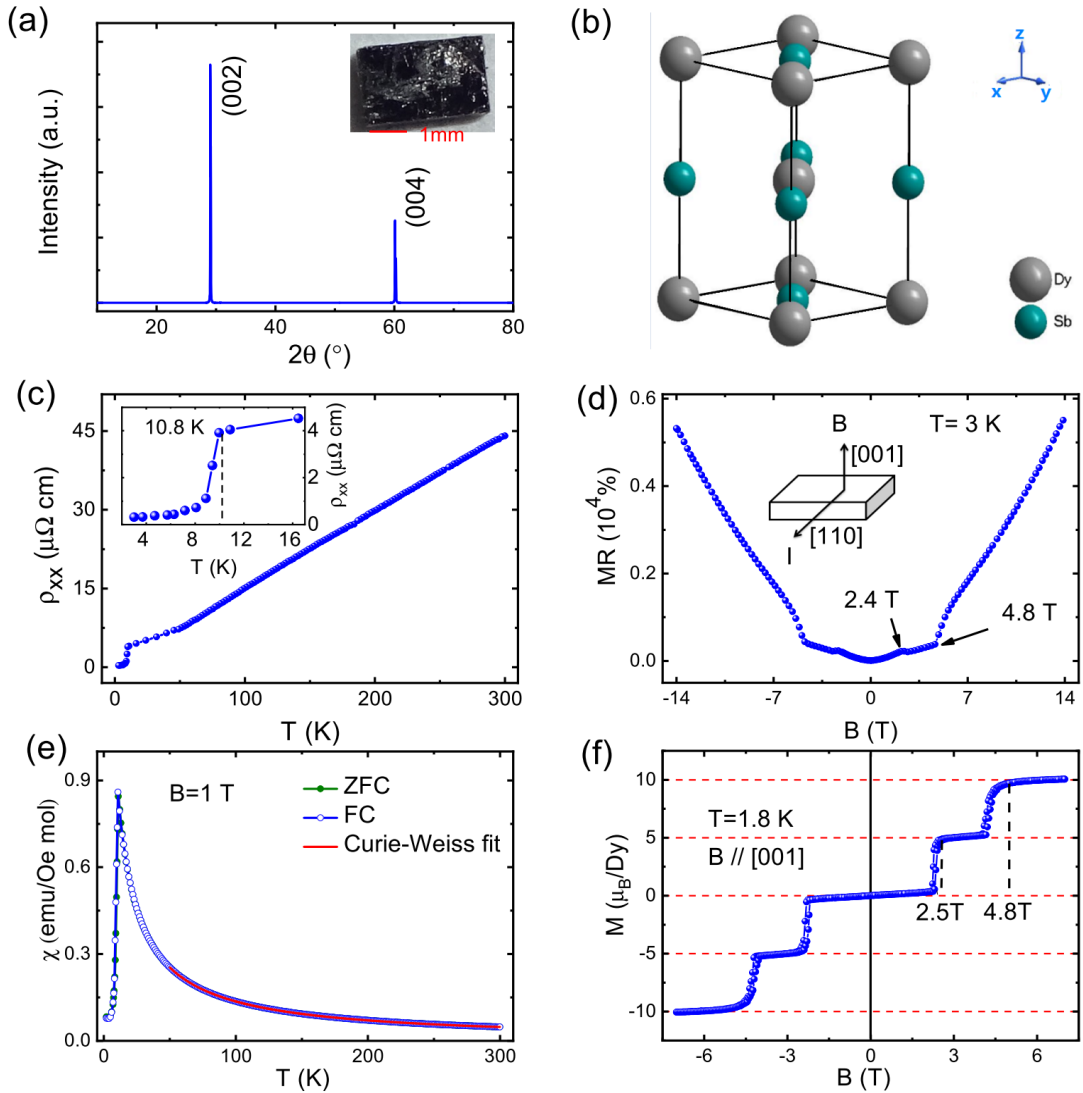

FIG. 2. (a) XRD pattern on the (00*l*) surface of single crystal DySb. Inset shows the as-grown single crystal. (b) Tetragonal crystal structure of DySb at low temperature. (c) Dependence of resistivity $\rho_{xx}$ on temperature from 3 K to 300 K. Inset: enlarged view for low temperature. (d) MR=$\frac{\rho(B)-\rho(0)}{\rho(0)}$ taken at 3 K. Inset illustrates the measurement geometry. (e) Temperature dependent susceptibility (zero field cooling, ZFC and field cooling, FC) measured in a field of 1 T. Red solid line represents the Curie-Weiss fit to the data above 50 K. (f) Magnetic hysteresis loop taken at 1.8 K up to 7 T.

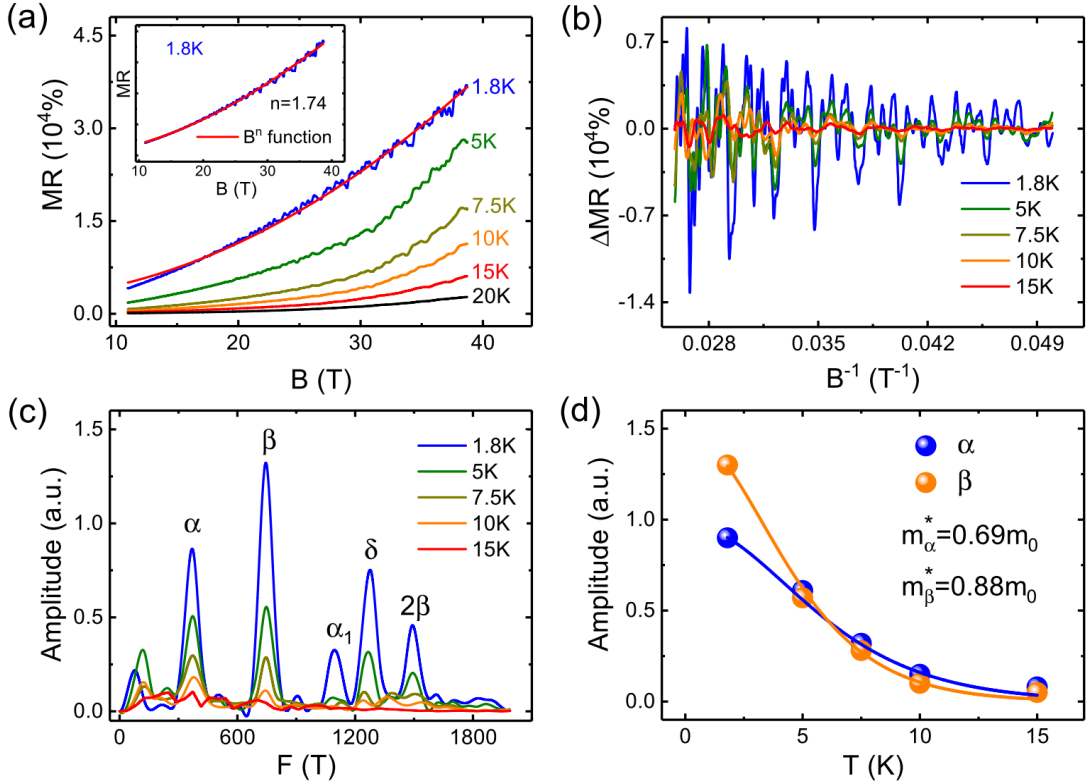

FIG. 3. (a) High field MR taken at various temperatures in the field range of 11-38.7 T. Red solid curve represents the fit to Lorentz law, i.e., MR=$(\mu_{ave}B)^2$. Inset: MR at 1.8 K and the power law fit (red solid curve). (b) Oscillating part of MR as a function of $B^{-1}$. (c) FFT spectra of SdH oscillations in (b). (d) Frequency amplitudes plotted against temperature for α and β. Solid lines are the fits using the amplitude expression of Lifshitz-Kosevich formula.

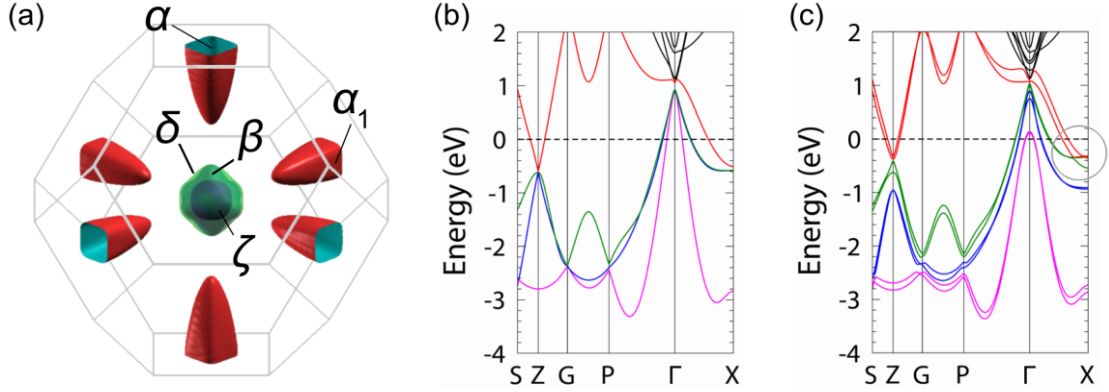

FIG. 4. (a) Calculated Fermi surface in the first Brillouin zone for DySb. Spin polarization is set as FM. (b) Band structures without SOC, plotted along high symmetry points of BZ. (c) Band structures with SOC considered. Only spin up is shown, as spin up and spin down are the same for this plot. Grey circle highlights a band inversion along ΓX.

TABLE I. Parameters obtained from the FFT analyses of SdH oscillations at $T=1.8$ K and from the DFT calculations.

|  | α | β | $α_1$ | δ | 2β |
|---|---|---|---|---|---|
| $F^{exp}$ (T) | 369 | 745 | 1094 | 1276 | 1492 |
| $A$ (nm$^{-2}$) | 3.52 | 7.10 | 10.43 | 12.16 |  |
| $k_F$ (nm$^{-1}$) | 1.06 | 1.50 | 3.13 | 1.97 |  |
| $n$ (cm$^{-3}$) | $3.57 \times 10^{20}$ | $1.14 \times 10^{20}$ |  | $2.58 \times 10^{20}$ |  |
| $F^{cal}$ (T) | 369 | 1022 | 1063 | 1206 |  |
| $m^{*cal}$ ($m_0$) | 0.21 | 0.42 | 0.49 | 0.48 |  |
| carrier type | electron | hole | electron | hole |  |